\RequirePackage{lineno}
\documentclass[aps,prl,superscriptaddress,showpacs,twocolumn]{revtex4-1}
\usepackage{graphicx}
\usepackage{floatflt}
\usepackage{float}
\usepackage{tabularx}
\usepackage{comment}
\usepackage{mathtools}

\usepackage{amssymb}

\begin{document}



\title{Internal Consistency of Neutron Coherent Scattering Length Measurements from Neutron Interferometry and from Neutron Gravity Reflectometry}
\author{W. M. Snow}
\email[Corresponding author: ]{wsnow@indiana.edu}
\author{J. Apanavicius}
\author{K. A. Dickerson}
\author{J. S. Devaney}
\author{H. Drabek}
\author {A. Reid}
\author {B. Shen}
\author{J. Woo}
\affiliation{Indiana University/CEEM, 2401 Milo B. Sampson Lane, Bloomington, IN 47408, USA}
\author{C. Haddock}
\affiliation{Center for Neutron Research, National Institute for Standards and Technology, Gaithersburg, MD 20899, USA }
\author{E. Alexeev}
\affiliation{University of California, San Diego, La Jolla, CA 92093}
\author{M. Peters}
\affiliation{Massachusetts Institute of Technology, Cambridge, MA 02138}

\date{\today}

\begin{abstract}
Many theories beyond the Standard Model postulate short-range modifications to gravity which produce deviations of Newton's gravitational potential from a strict $1/r$ dependence. It is common to analyze experiments searching for these modifications using a potential of the form $V^{\prime}(r)=-\frac{GMm}{r} [1+\alpha \exp{(-r/\lambda)}]$. The best present constraints on $\alpha$ for $\lambda <100$\,nm come from neutron scattering and often employ comparisons of different measurements of the coherent neutron scattering amplitudes $b$. We analyze the internal consistency of existing data from two different types of measurements of low energy neutron scattering amplitudes: neutron interferometry, which involves squared momentum transfers $q^{2}=0$, and neutron gravity reflectometry, which involves squared momentum transfers $q^{2}=8mV_{opt}$ where $m$ is the neutron mass and $V_{opt}$ is the neutron optical potential of the medium. We show that the fractional difference $\frac{\Delta b}{|b|}$ averaged over the 7 elements where high precision data exists on the same material from both measurement methods is $[2.2 \pm 1.4] \times 10^{-4}$. We also show that $\frac{\Delta b}{|b|}$ for this data is insensitive both to exotic Yukawa interactions and also to the electromagnetic neutron-atom interactions proportional to the neutron-electron scattering length $b_{ne}$ and the neutron polarizability scattering amplitude $b_{pol}$. This result will be useful in any future global analyses of neutron scattering data to determine $b_{ne}$ and bound $\alpha$ and $\lambda$. We also discuss how various neutron interferometric and scattering techniques with cold and ultracold neutrons can be used to improve the precision of $b$ measurements and make some specific proposals.   

\end{abstract}



\maketitle

\section{Introduction and Theoretical Overview}

Newton's inverse square law form for the force of gravity between two point-like test bodies is one of the first quantitative facts learned by students of physics. In the limit where relativistic effects are negligible, this law is obeyed with high accuracy over macroscopic distance scales.
Many theoretical speculations, however, propose that the $1/r^{2}$ gravitational force law can be greatly modified at shorter distances. Examples of these speculations include the idea of compact extra dimensions of spacetime accessible only to the gravitational field, which can explain the unnaturally small strength of gravity relative to the other known forces~\cite{Arkani98, Adelberger03, Frank2004} and the idea that gravity might be modified on the length scale of 100 microns corresponding to the scale set by the dark energy density~\cite{Adelberger09}. Experiments which search for possible modifications to gravity at short range are also sensitive to new non-gravitational interactions of various types. Many extensions to the Standard Model of particle physics produce weakly-coupled, long-range interactions~\cite{Jae10, Antoniadis11}. Certain candidates for dark matter in the sub-GeV mass range can induce Casimir-Polder-type interactions between nucleons~\cite{Fichet2017, Brax2018} with ranges from nuclear to atomic scales. New sources of information which can probe exotic gravity or other possible exotic interactions on short distance scales are therefore of fundamental interest. 

Many experiments have been conducted to search for short-range deviations from the $1/r^{2}$ gravitational force law~\cite{Adelberger09, Murata15}. Most of the results from experimental searches have been analyzed assuming a potential of the form

\begin{equation}
\label{eq:yukawa:gravity}
V^{\prime}(r)=-\frac{GMm}{r} [1+\alpha \exp{(-r/\lambda)}]
\end{equation}


\noindent where $G$ is the gravitational constant, $m_{1,2}$ are the masses of two objects separated by a distance $r$, and  $\alpha$ and $\lambda$ parametrize the strength of some new Yukawa interaction relative to gravity and the range set by the mass of the new massive boson whose exchange generates the new potential.
Recent reviews~\cite{Adelberger09, Murata15} present the existing limits on $\alpha$ and $\lambda$, which come from torsion balances~\cite{Kapner07}, and microcantilevers and techniques adapted from measurements of the Casimir effect~\cite{Mohideen98, Sushkov11, Chiaverini03, Chen14}. Experiments using laser-levitated dielectric microspheres~\cite{Geraci10, Rider2016} are also in progress.

Below 100 nanometers the most stringent experimental limits for many of these weakly-coupled exotic interactions come from experiments using neutrons. The electrical neutrality of the neutron coupled with its small magnetic moment and very small electric polarizability make it insensitive to many of the electromagnetic backgrounds such as the Casimir effect which can plague experiments that employ test mass pairs made of atoms. The ability of slow neutrons to penetrate macroscopic amounts of matter and to interact coherently with the medium allow the quantum amplitudes governing their motion to accumulate large phase shifts which can be sensed with interferometric measurements~\cite{Nico05b, Dubbers11, Pignol:2015}. These features of slow neutron interactions with matter and external fields have been exploited in a number of recent experiments which search for possible new weakly-coupled interactions of various types~\cite{Leeb92, Bae07, Ser09, Ig09, Pie12, Yan13, Lehnert14, Lehnert15, Jen14, Lemmel2015, Li2016, Lehnert2017, Haddock2018b, Cronenberg2018}. This strategy can succeed despite the uncertainties in our knowledge of the neutron-nucleus strong interaction. In the slow neutron regime with $kR\ll1$ where $k$ is the neutron wave vector and $R$ is the range of the neutron-nucleus strong interaction, neutron-nucleus scattering amplitudes are dominated by s-wave scattering lengths which are accurately measured experimentally. This makes coherent neutron interactions with matter sufficiently insensitive to the complicated details of the strong nucleon-nucleus interaction that one can cleanly interpret and analyze searches for smaller effects. Existing neutron limits on deviations from the $1/r^{2}$ gravitational force law between $10^{-8}-10^{-12}$ m come from theoretical analyses of the neutron energy and $A$ dependence of neutron-nucleus scattering lengths~\cite{Nez08}, which have been measured to better than $0.1$\% accuracy for a large number of nuclei. Other experiments have measured the angular distribution of neutrons scattered from noble gases to search for a deviation from that expected in this theoretically-calculable system~\cite{Kamiya2015, Haddock2018a}. At shorter distances the best limits come from the measured energy dependence of neutron-nucleus cross sections in lead~\cite{Leeb92, Pokot06} and from very high energy forward cross section measurements at accelerator facilities~\cite{Kamyshkov08}. 

Various authors~\cite{Leeb92, Zimmer2006, Nez08} have conducted analyses of the neutron scattering data to constrain exotic Yukawa interactions. All used some amount of theoretical modeling of the neutron Standard Model interactions in combination with experimental information. In all of these cases uncertainties in the neutron-atom strong and electromagnetic interactions still place an ultimate limit on the sensitivity of these types of searches for possible new interactions. For the case of slow, unpolarized neutrons incident upon unpolarized atoms with energies far from neutron-nucleus resonances, the s-wave neutron-atom scattering amplitude $b_\mathrm{atom}(q)$ as a function of the momentum transfer $q$ can be expressed as~\cite{Sears1986}

\begin{equation}
\label{batom}
b_\mathrm{atom}(q) = b(q)-b_{ne}Z[1-f(q)] +b_{pol}(q)+b_{Y}(q)\,.
 \end{equation}

The first term $b(q)$ describes the low energy (s-wave) coherent neutron scattering from the nucleus of the atom from the neutron-nucleus strong interaction. The second term describes the interaction between the internal radial charge density of the neutron and the electric field of the atom. It is proportional to the neutron electron scattering length $b_{ne} = -1.345(25) \times 10^{-3}$ fm~\cite{Tan18} and depends on the atomic form factor $f(q)$ of the electron distribution around a nucleus of charge Z, which is measured by x-ray scattering. The third term $b_\mathrm{pol}(q)$ is proportional to the very small but nonzero electric polarizability of the neutron and comes from the neutron electric dipole moment induced by the very large electric field near and inside the nucleus. Finally the forth term $b_{Y}(q)=-f_{Y}(q)={2G \alpha m^{3}A \over \hbar^{2}}{1 \over (q^{2}+1/\lambda^{2})}$ comes from applying the Born approximation to calculate the scattering amplitude corresponding to the exotic Yukawa interaction potential of interest in this work and from adopting the convention $b=-f$ historically used in slow neutron scattering.

In this paper we will analyze the possible effects of both exotic Yukawa interactions and of electromagnetic neutron-atom interactions in the context of Eqn. 2.  The different $q$'s used in various neutron-atom scattering amplitude experiments weight the contributions in this expression differently. In principle one needs to perform some type of global analysis of the scattering data~\cite{Leeb92, Zimmer2006, Nesvizhevsky2006, Nez08} to derive constraints on $b_{Y}(q)$. In the near future we expect that new more sensitive data will be available which can enable an improved analysis. Such a new analysis is beyond the scope of this paper. For such an improved analysis, however, it would be very useful to investigate the degree of internal consistency in the existing data set on coherent neutron scattering amplitudes. In this paper we show that we can test the internal consistency of $b_{atom}(q)$ measurements using slow neutrons in an essentially model-independent way using data from the two most sensitive neutron optical techniques. Forward scattering techniques for $b_{atom}(q)$ measurement such as neutron transmission and neutron interferometry involve squared momentum transfers $q^{2}=0$, and neutron gravity reflectometry involves squared momentum transfers $q^{2}=8mV_{opt}$ where $m$ is the neutron mass and $V_{opt}$ is the optical potential of the medium. The precision of both of these methods approaches the $10^{-4}$ level. Of the many atomic species which have been measured to high precision by these techniques, there is a subset where data exists for the same medium from both techniques. Furthermore, over the small range of squared momentum transfers $0< q^{2}< 8mV_{opt}$ and over the meV energy range spanned by these two measurement methods, the differences in the neutron-electron interactions term $b_{ne}Z[1-f(q)]$, the neutron polarizability term $b_{pol}(q)$, and the exotic Yukawa term $b_{Y}(q)$ are all at least three orders of magnitude smaller than the present experimental uncertainties in $b_{atom}(q)$. Therefore the fractional difference ${(b_{GR}-b_{T}) \over |b_{T}|}={\Delta b \over |b|}$,  where $b_{GR}$ comes from gravity reflectometry and $b_{T}$ comes from neutron interferometry,  can be used to judge the internal consistency of the neutron scattering amplitude data set independently of one's knowledge of the neutron electromagnetic and exotic gravity interactions. The simple geometries and macroscopic sample sizes used in neutron interferometry and neutron gravity reflectometry experiments which we analyze, combined with the availability of analytic solutions to the effects of a weak perturbation of Yukawa form on these observables, makes it possible to evaluate the different corrections from a Yukawa interaction analytically to high accuracy for each case. At this level of precision one must also take into account some small multiple scattering corrections to the kinematic limit of neutron optics whose physical origin we briefly review below. 

The result of our analysis is quite encouraging. We find that $\frac{\Delta b}{|b|}$ for the 7 nuclei which have been precisely measured using both techniques is consistent with zero at the $10^{-4}$ level. This result demonstrates the internal consistency of the associated data and can be applied to analyses of neutron data searching for exotic Yukawa interactions from future experiments. We mention some of the issues that must be carefully considered in any future global neutron scattering analysis to constrain exotic interactions which makes use of a wider dynamic range of neutron energies and momentum transfers. We also outline how the sensitivity of this approach to constraining exotic Yukawa interactions can be improved by about $1-2$ orders of magnitude through future higher-precision coherent neutron scattering length measurements using neutron interferometry for $b_\mathrm{T}$ combined with future high-precision measurements using ultracold neutrons (UCN) for $b_\mathrm{GR}$.

The rest of this paper is organized as follows. We first present the expressions for the modification of the neutron optical potential and the neutron interferometer phase shift from a slab of matter in the presence of an extra Yukawa interaction. Next we present the correction to the neutron optical reflectivity if one adds an exotic Yukawa potential to the neutron-atom interaction. We estimate the size of the difference between $b_{GR}$ and $b_{T}$ from the electromagnetic and Yukawa terms. We gather the neutron scattering length data for nuclei which have been measured by both techniques with high precision and analyze this data to demonstrate their degree of internal consistency. We end by outlining additional neutron interferometry measurements which can be compared to the existing neutron gravity reflectometry measurements and outline future measurements using cold and ultracold neutrons.

\section{Corrections to the Phase Shifts measured in Neutron Interferometry from a Weak Yukawa potential}
\label{sec:transmission:methods}

The most sensitive methods for the measurement of forward neutron scattering amplitudes comes from perfect crystal neutron interferometry, which is described in great detail in a recent work~\cite{RauchWerner}. Neutron interferometric measurements of scattering amplitudes employ a Mach-Zehnder interferometer in which the neutron amplitude $\psi e^{-i\Phi}$ is coherently split into two paths and recombined using perfect crystal dynamical diffraction. The measured phase shift is dominated by the real part of the neutron optical potential $V(x)$ and can be expressed as~\cite{RauchWerner}

\begin{equation}
    \label{eq:Normalphaseshift}
    \Phi={m \over k \hbar^{2}}\int{V(x)dx}
\end{equation}

\noindent where $m$ is the neutron mass and $k$ is the neutron wave number. For $kR\ll1$ where $R$ is the range of the neutron-atom interaction, $V(x)=V_{F}$ where $V_{F}$ is the Fermi pseudopotential from the short-range strong and electromagnetic interactions of the neutron with the atoms in the material. In the kinematic limit of the theory of neutron optics, $b$ is related to $V_{F}$ by

\begin{equation}
    \label{eq:FermiPotential}
    V_{F}={2\pi\hbar^{2} N b \over m}
\end{equation}



\noindent where $N$ is the atom number density. The sample geometry used in all neutron interferometry scattering length measurements employs a rectangular plate of thickness $L$ whose surface is normal to one of the coherent subbeams in the interferometer. In the presence of an additional Yukawa interaction between the neutron and the sample material with a range $L\gg\lambda>R_{atom}$ we must integrate the accumulated phase shift from the potential from a plate of matter of uniform mass density $\rho$. Since the thickness of the samples is much greater than the range $\lambda$ of the Yukawa interaction, and the neutron transverse coherence length is very small compared to the transverse dimensions of the samples in all of the neutron interferometry scattering length measurements, the potential energy of a neutron as a function of $x$, the distance from the neutron to the plate, can be calculated analytically for a Yukawa potential by taking the limit of an infinite planar slab of material as~\cite{Zimmer2006, Greene2007}

\begin{equation}
    \label{eq:Yukawaopticalout}
  V(x)=-V_\mathrm{Y}\exp{(-|x|/\lambda)}
\end{equation}
outside the material, and 
\begin{equation}
    \label{eq:Yukawaopticalin}
  V(x)= V_{F}-V_\mathrm{Y}[2-\exp{(-|x|/\lambda)}]
\end{equation}
inside the material, where $V_\mathrm{Y} = -2Gm\pi\rho\alpha\lambda^{2}$. By treating the exotic Yukawa interaction as a weak perturbation compared to $V_{F}$, the additional neutron phase shift from the Yukawa interaction can be calculated as~\cite{Greene2007}

\begin{equation}
    \label{eq:Yukawaphaseshift}
 \Delta \Phi_\mathrm{Y}=\frac{-2mV_\mathrm{Y} (L+2\lambda) }{\hbar^{2}k}
\end{equation}

where we have approximated $\rho=Nm$. By comparison of eqs.
\ref{eq:Normalphaseshift}, \ref{eq:FermiPotential}, and \ref{eq:Yukawaphaseshift} we can express the effect of the Yukawa-like deviation from gravity for the case of neutron interferometry in terms of an additional contribution to the coherent scattering amplitude $b_{T}$. One can split the effect of the Yukawa interaction shown in eq.~\ref{eq:Yukawaphaseshift} into a ``bulk" term, which just adds to the Fermi potential $V_{F}$, and a term from the tail of the Yukawa potential which extends outside the slab on both ends. From eq.~\ref{eq:Yukawaphaseshift} one can see that the size of the ``tail" term for interferometry is smaller than the bulk Yukawa term by a factor of $2\lambda/L$ where $L$ is the sample thickness, of order 1\,mm or greater in neutron interferometry measurements. Therefore for the range of $\lambda<100$\,nm of interest for neutron constraints on exotic Yukawa interactions $\lambda/L<10^{-4}$ and the dominant correction term for $b_{T}$ becomes

\begin{equation}
    \label{eq:byNI}
b_{T} = -\frac{2 \alpha G m^{3}A\lambda^{2}}{\hbar^2}.
\end{equation}

\begin{table*}[t]
\begin{center}
\begin{tabular}{llllllllll}
\hline
Element & A & $b_\mathrm{GR}$ & $\delta b_\mathrm{GR}$ & Refs.(GR) & $b_{T}$ & $\delta b_{T}$ & Refs.(T) & $\Delta b/b_{T}$ & $\delta [\Delta b/b_{T}]$    \\
 $^{1}$H & 1 & -3.7406  & 0.0011 & \cite{Nistler1974, Koester1975, Sears1985b} &  -3.7384  & 0.0020 & \cite{Schoen2003} & -0.00058 & 0.00061  \\  
 $^{2}$H & 2 & 6.6713  & 0.0036 & \cite{Nistler1974, Sears1985b} &  6.6649   & 0.004 & \cite{Schoen2003} & 0.00096 & 0.00081  \\ 
C & 12 & 6.6460  & 0.0012 & \cite{Koester1975, Sears1985b} &  6.6484   & 0.0013 &\cite{Koester1979} & -0.00015 & 0.00028  \\ 
O & 16 & 5.8025  & 0.0041 & \cite{Nistler1974, Sears1985b} &  5.805   & 0.004 & \cite{Nistler1974, Koester1979}& -0.00043 & 0.00099  \\ 
Sn & 119 & 6.2257  & 0.0019 & \cite{Reiner1990} & 6.2220   &  0.0018 & \cite{Bauspiess1978}  & 0.00092 & 0.00044   \\ 
  Pb & 207 & 9.4031  & 0.0015 &\cite{Reiner1990, Koester1986, Sears1985b} &  9.4017   & 0.002 & \cite{Ioffe2000}& 0.00015 & 0.00022  \\ 
  Bi & 209 & 8.5284  & 0.0011 & \cite{Nucker1969, Reiner1990, Sears1985b} &  8.5201  & 0.0034 &\cite{Rauch1987, Tuppinger1988} & 0.00097 & 0.00042  \\          
 
\hline
\end{tabular}
\caption{A list of the neutron-nucleus scattering length measurements used in this analysis which have been conducted using the techniques described above. All scattering length units are in fm. The measurements using the gravity reflectometry method $b_{GR}$ were all performed at the FRM research reactor by the group of Koester {\it et al.} The measurements of $b_{T}$ all come from neutron interferometry. The $b_{T}$ values for H and D come from the analysis presented in the appendix of Schoen {\it et al.}~\cite{Schoen2003}. The scattering length value for C comes from two separate neutron interferometer measurements of $^{12}$C and $^{13}$C properly weighted in order to be able to compare to the results from Koester, which used liquids with natural isotopic abundance. The $b_{GR}$ values and the $b_{T}$ values include small corrections for neutron optics multiple scattering effects as evaluated by Sears~\cite{Sears1985b} and Schoen~\cite{Schoen2003}. All of the interferometer measurements except for Schoen {\it et al.} were conducted at high enough neutron energies that these multiple scattering corrections are negligible. The accuracy for the scattering amplitudes achieved by both techniques is comparable.}
\label{tbl:lengths}
\end{center}
\end{table*}  

\section{Corrections to the Reflectivity measured in Neutron Gravity Reflectometry from a Weak Yukawa potential}
\label{sec:gravity:reflectometry}

The gravity reflectometry method for the measurement of scattering amplitudes~\cite{Leibnitz1962, Koester1965} has also produced n-A scattering amplitude results of high precision. In this method one prepares a slow neutron beam which drop in the gravitational field of the Earth by a height $H$ over a long evacuated flight path so that all of the neutrons in the beam gain an extra momentum along the gravitational field corresponding to an energy $E=mgH$. This neutron beam is allowed to fall upon a flat mirror made of the material of interest of neutron optical potential $V_{F}$,  which is maintained in liquid form so that the surface is normal to the direction of the local gravitational field. When $E=V_{F}$ the neutrons start to penetrate the mirror and the reflectivity $|R|^{2}$ falls below unity according to the well-known Fresnel reflectivity formula of optics. A precise measurement of $|R|^{2}$ as a function of $H$ can determine $V_{F}$ and therefore the n-A scattering amplitude $b_{GR}$. A long series of such measurements on many materials spanning nearly three decades was conducted by the group of Koester {\it et al.} at the FRM research reactor in Garching, Germany on a specialized neutron beamline devoted specifically for this purpose. All of the reflectometry data analyzed in this paper comes from this group. 




We show below that, to high accuracy in our regime of interest, the presence of a neutron-atom Yukawa potential just modifies the arguments in the Fresnel reflectivity expression used to analyze this data while preserving its functional form. The modification to the formula for the reflectivity for an exponential potential can be calculated~\cite{Taketani2012} with the exotic Yukawa potential treated as a perturbation as also done for the neutron interferometry case above. In the limit where the neutron mirror is treated as an infinite plane, the 1D Schr{\"o}dinger wave equation for this potential can be solved exactly. The solutions are proportional to the modified Bessel functions of the first kind. The reflection amplitude $R$ of a neutron incident upon the surface of the material is given by demanding continuity of the wave function and its logarithmic derivative at the surface

\begin{equation}
  R = -\frac{\phi_{o,+}(0)}{\phi_{o,-}(0)}
  \frac{\frac{d}{dz}\text{ln}\phi_{i,+}(z) - \frac{d}{dz}\text{ln}\phi_{o,+}(z)}{\frac{d}{dz}\text{ln}\phi_{i,+}(z) - \frac{d}{dz}\text{ln}\phi_{o,-}(z)}\bigg\rvert_{z=0}
\end{equation}

\noindent where $\phi_{o,+}$ ($\phi_{i,+}$) and $\phi_{o,-}$ ($\phi_{i,-}$) are the two independent solutions to the wave equation outside (inside) of the material. Since the wave equation is of second order and we are considering two independent cases, we must have four independent solutions in total. In the limit where the neutron interaction energy with the mirror from the Yukawa interaction is much smaller than the kinetic energy of the incident neutron, and we restrict ourselves to the regime of ($\alpha$-$\lambda$) parameter space of interest in this work and where this reflectivity calculation is relevant, namely $\lambda=(1-100)\times 10^{-10}$ m  and $\alpha$ below the existing experimental limits the expression for the reflection probability $|R|^{2}$ can be written as (see the Appendix)

\begin{equation}
  |R|^2= \left[\frac{1-\sqrt{1-H_c^{\prime}/H}}{1+\sqrt{1-H_c^{\prime}/H}}\right]^2
\end{equation}

\noindent where $  H_c^{\prime} = \frac{V_\mathrm{F}+2V_\mathrm{Y}}{mg}$ is the critical height in the presence of a new Yukawa deviation from gravity proportional to $V_\mathrm{Y}$, and $H_c^\prime < H$. 

 This is the same Fresnel reflectivity formula used by the Koester group to analyze their data, but with the critical height $H_c$ replaced with $H_c^{\prime}$. The value of the critical height is determined experimentally when the reflectivity curve becomes discontinuous and $H_c^\prime = H$. In the presence of the Yukawa interaction, this height shift can be expressed in terms of scattering lengths as

\begin{equation}
  b^{\prime}=\frac{m}{ 2\pi N \hbar^{2}}\times[mgH_{c}+2V_\mathrm{Y}]=b+\frac{mV_\mathrm{Y}}{ \pi N \hbar^{2}}
  \label{eq:new_bg}
\end{equation}

\noindent where $b$ is the coherent scattering length inferred from the gravity reflectometry data for the case of no Yukawa interaction and $b^{\prime}$ is the scattering amplitude in the presence of the Yukawa interaction. We can therefore identify

\begin{equation}
  b_{GR} = \frac{mV_\mathrm{Y}} {\pi N \hbar^{2}} = -\frac{2\,\alpha\,G\,m^{3} A\lambda^{2}}{\hbar^2}
  \label{eq:new_bg:Y}
\end{equation}

and by comparing equations \ref{eq:byNI} and \ref{eq:new_bg:Y} we see that $b_{Y, GR}-b_{Y, T}=0$ to high accuracy. This can be understood simply. In both cases one can split the effect of the Yukawa interaction on the observable of interest into a ``bulk" term which just adds to the the Fermi potential $V_{F}$ and a term from the ``tail" of the Yukawa potential which extends outside the slab. As the bulk term is the same for both cases it cancels in the difference. The size of the tail terms for interferometry and gravity reflectometry are not exactly the same, but they are both much smaller than the bulk Yukawa term by a factor below $10^{-3}$ in both cases for the range of $\lambda<100 $\,nm of interest for neutron constraints on exotic Yukawa interactions. 


\section{Corrections to n-A scattering Lengths from Neutron-Atom Electromagnetic Interactions}
\label{sec:corrections:electromagnetic}

The remaining sources for the difference $\Delta b=[(b_\mathrm{atom,GR}-b_\mathrm{atom,T}]$ between the scattering lengths measured by these two different methods come from Standard Model interactions. For the case of slow, unpolarized neutrons incident upon unpolarized atoms, $b_\mathrm{atom}(q)$ can be expressed as~\cite{Sears1986}

\begin{equation}
\label{batom}
b_\mathrm{atom}(q) = b(q)-b_{ne}Z[1-f(q)] +b_{pol}(q)\,.
 \end{equation}

The first term $b$ describes the low energy (s-wave) scattering from the nucleus of the atom from the neutron-nucleus strong interaction, which has contribution from both the potential scattering and (for heavier nuclei) from the low-energy tails from n-A resonance scattering. The resonances contribute to a slight dependence of $b(q)=b_{pot}+b_{res}$ on neutron energy through the Breit-Wigner resonance formula. In the presence of n-A resonances the expression for the resonant part $b_{res}$ of the total scattering amplitude  becomes~\cite{Mughabghab81}

\begin{equation}
     b_{res}=\sum_{j}{g_{\pm, j} \over 2k^{'}_{j}}{\Gamma_{n,j} \over [(E^{'}-E_{j})+i\Gamma_{j}/2]}
\end{equation}

\noindent where $\Gamma_{n,j}$ and $\Gamma_{j}$ are the neutron width and total width of the resonance at energy $E_{j}$ and $k^{'}=\mu k/m$ is the wave vector in the n-A center of mass system of reduced mass $\mu$,  $E^{'}$ is the associated energy in the COM frame, and $g_{+, j}=(I+1)/(2I+1)$ and $g_{-,j}=I/(2I+1)$ are the statistical weight factors for a resonance at energy $E_{j}$ in the total angular momentum channels $J=I \pm 1/2$. This means that the neutron scattering amplitudes that are reported in the literature from slow neutron measurements are in fact a sum of the potential scattering contribution and also the tails of all of the other resonances in the limit $E \to 0$: 

\begin{equation}
     \label{eq:resonancetails}
 b_{measured}=R-\sum_{j}{g_{\pm, j} \over 2k^{'}_{j}}{\Gamma_{n,j} \over [(E_{j})-i\Gamma_{j}/2]}
 \end{equation} 

\noindent and since $\Gamma$ scales linearly with $k^{`}$, this expression gives a finite contribution in the $k^{`} \to 0$ limit. 

The second term $b_{ne}Z(1-f(q))$ describes the interaction between the internal radial charge density of the neutron and the electric field of the atom. It is proportional to the neutron electron scattering length $b_{ne} = -1.345(25) \times 10^{-3}$ fm~\cite{Tan18} and depends on the atomic form factor $f(q)$ of the electron distribution around a nucleus of charge Z, which is measured by x-ray scattering and obeys approximately the universal form $f(q) = \frac{1}{\sqrt{1+3(q/q_x)^2}}$ where the element-specific parameter $q_x$ of order $\frac{\hbar}{R_\mathrm{atom}}$ can be obtained from fitting to the x-ray scattering data. The corresponding form factor from the internal charge distribution of the nucleus can be expanded in the small $q$ limit as $F(q)=1-{1\over 6} (qR')^{2}$ where $R'$ is the root mean square nuclear charge radius~\cite{Hofstader1956}. The third term $b_\mathrm{pol}(q)$ is proportional to the very small but nonzero electric polarizability of the neutron and comes from the neutron electric dipole moment induced by the very large electric field near and inside the nucleus~\cite{Thaler1959, Leeb1984, Sears1986, Schmied1988}. In the small $q$ limit $b_{p}={{Z^{2}e^{2}m\alpha_{em}} \over {\hbar^{2}R}}[6/5-\pi qR/4]$ where $Ze$ is the nuclear charge, $\alpha_{em}$ is the electromagnetic coupling, and $R$ is the nuclear radius. The first term is $q$-independent and is as large as $0.06$ fm for uranium.

We can now estimate the size of $\Delta b=b_{GR}-b_{T}$ knowing the slightly different energy and momentum transfer ranges accessed in these measurements. The typical relative sizes of $b(q)$, $b_{ne}Z[1-f(q)]$, and $b_\mathrm{pol}(q)$ in the slow neutron regime for medium-mass nuclei are in the approximate proportion $1:10^{-2}:10^{-3}$.  The neutron interferometry measurements all possess $q^{2}=0$ and were conducted at neutron energies of several meV. The neutron gravity reflectometry measurements all possess $q^{2}=8mV_{opt}$ with q's below $10^{-2}$ inverse Angstroms, and the energies used on the Koester et al measurements were centered at $0.5$ meV. Using the expressions above we can see that the contribution to $\Delta b$ from the $b_{ne}$ term is of order $10^{-6}$ and that from the $b_{p}$ term is of order $10^{-15}$. The contribution from the energy dependence of the tails of the n-A resonances depends on the details of the resonance energies and widths of the particular nuclei. For the particular list of nuclei used in the analysis of this paper (H, D, C, O, Sn, Pb, and Bi) the light nuclei possess no n-A resonances, and both Pb and Bi are close in A to the doubly-magic nucleus $^{208}$Pb, which possess especially low level densities near threshold and in particular no low-lying resonances between $1-10$ eV whose tails could give a visible energy dependence in the meV regime. As for Sn: three of its isotopes have  n-A resonances between $0-10$ eV~\cite{NNDC,Shibata2011}: $^{113}$Sn ($E=8.3$ eV, $\Gamma_{n}=4.5$ meV), $^{117}$Sn ($E=1.3$ eV, $\Gamma_{n}=0.00011$ meV, a p-wave resonance), and $^{119}$Sn ($E=6.2$ eV, $\Gamma_{n}=0.00148$ meV). Using the real part of the resonance formula above one sees that the contributions to $\Delta b/|b|$ from the residual neutron energy dependence of $b_{res}$ for Sn over a $\delta E=10$ meV range starting at 0.5 meV is of order ${\Gamma_{n}\delta E} \over {E_{res}^{2}}$, which does not exceed $10^{-6}$ for any of these resonance parameters. This is much smaller than the precision of the $b$ measurements analyzed in this paper, which do not exceed $10^{-4}$. 

We conclude that all of the physical effects analyzed above, both from a possible exotic Yukawa interaction and from Standard Model interactions, give differences well below the current measurement precision for the scattering lengths. Neither exotic Yukawa interactions nor Standard Model neutron-atom interactions can introduce a visible difference between these two methods of neutron scattering length measurements. Therefore an analysis of $\Delta b/|b|$ from these two methods is a valid test of the internal consistency of the present experimental data.  





\section{Corrections to n-A scattering Lengths from Multiple Scattering Effects in the Neutron Optical Potential}
\label{sec:corrections:multiple:scattering}

Before comparing the scattering lengths determined by these two methods we must consider some small corrections to the usual kinematic expression for the neutron optical potential. The physical origin for these modifications comes from local field corrections and neutron multiple scattering in the medium and are physically very similar to the analogous corrections from dispersive effects for light optics in a dielectric medium. We make use of an evaluation of these effects performed long ago by Sears~\cite{Sears1985b}. His correction formulae are consistent with both previous and subsequent theoretical work using different theoretical approaches~\cite{Snow2019}. Calculations of multiple scattering corrections to the kinematic theory of neutron optics performed in the 80s~\cite{Sea82, Now82b,Die81,Now82a} built upon much earlier work~\cite{Foldy1945, Lax1951, Ekstein1951, Lax1952, Ekstein1953, Gol64, Lenk1975, Blaudeck1976} and were conducted within the framework of the traditional multiple scattering theory. Different calculational methods based on resummation of dominant subclasses of diagrams important for backscattering~\cite{Warner1985} and a Lindblad operator treatment developed to understand decoherence in neutron optics~\cite{Lanz1997} give the same results. All calculations restore consistency with the optical theorem and reduce in appropriate limits to the usual kinematic limit.

As shown by Sears, in the $kR\ll 1$ limit of relevance to this work the dominant correction to the neutron optical potential can be written in terms of a modified neutron index of refraction $n^{\prime}$:
\begin{equation}
     \label{eq:refractionindexSears:2}
     n^{\prime}=1-\frac{{2\pi N b^{\prime}} }{ {k^{2}}}[1+J^{\prime}+\frac{{\pi N b^{\prime}} }{k^2}]
\end{equation}
\noindent where the first two terms are the usual results from the kinematic theory of neutron optics and the last two terms come from local field effects and multiple scattering. $J^{\prime}=N b \int \exp{i{\vec k} \cdot {\vec{r}}} G(r) [1-g(r)] d^{3}r$ for an isotropic medium, where $G(r)=\exp{ikr}/r$ is the neutron Green's function and $g(r)$ is the pair correlation function for the atoms in the material, $n^{\prime}$ is the real part of the neutron index of refraction with the multiple scattering correction, $b^{\prime}$ is the neutron scattering length with the multiple scattering correction, $N$ is the number density of atoms in the material, and $k$ is the incident neutron wave vector. The neutron index of refraction is defined in the usual way by $n=k_{in}/k_{out}$ where $k_{in}$ and $k_{out}$ are the neutron wave vectors inside and outside of the medium. Sears shows that the resulting relationships between the true real part of the scattering length $b^{\prime}$ and the effective scattering length $b_\mathrm{eff}$ inferred foregoing the multiple scattering corrections is $b_\mathrm{T}=b^{\prime}[1+J^{\prime}+\frac{{\pi\rho b^{\prime}} }{ k^2}]$ for interferometry and $b_\mathrm{GR}=b^{\prime}[1+J^{\prime}]$ for gravity reflectometry. The correction for the neutron interferometry measurements performed using thermal neutrons (in the \(kb\ll 1\) limit) was found by Sears to be of order $10^{-5}$, more than one order of magnitude smaller than the measured accuracy and therefore negligible. For the cold neutron energies employed in the gravity reflectometry work the size of the corrections is at the $10^{-4}$ level, close to the size of some of the measurement errors for $b$.  Sears's scattering corrections were evaluated with $10$\% accuracy, which is about one order of magnitude more accurate than the scattering length measurement errors and therefore good enough for our analysis. Thus the precision with which these two methods of scattering length measurement can be compared is not yet limited by our knowledge of the multiple scattering corrections to the kinematic theory of neutron optics, a fact that will guide how this method to search for exotic short-range gravitational interactions of Yukawa form might be improved in the future.

\section{Data Evaluation and Analysis}
\label{sec:data:evaluation:analysis}

\begin{figure}[htb]
\centering
    \includegraphics[width=0.5\textwidth]{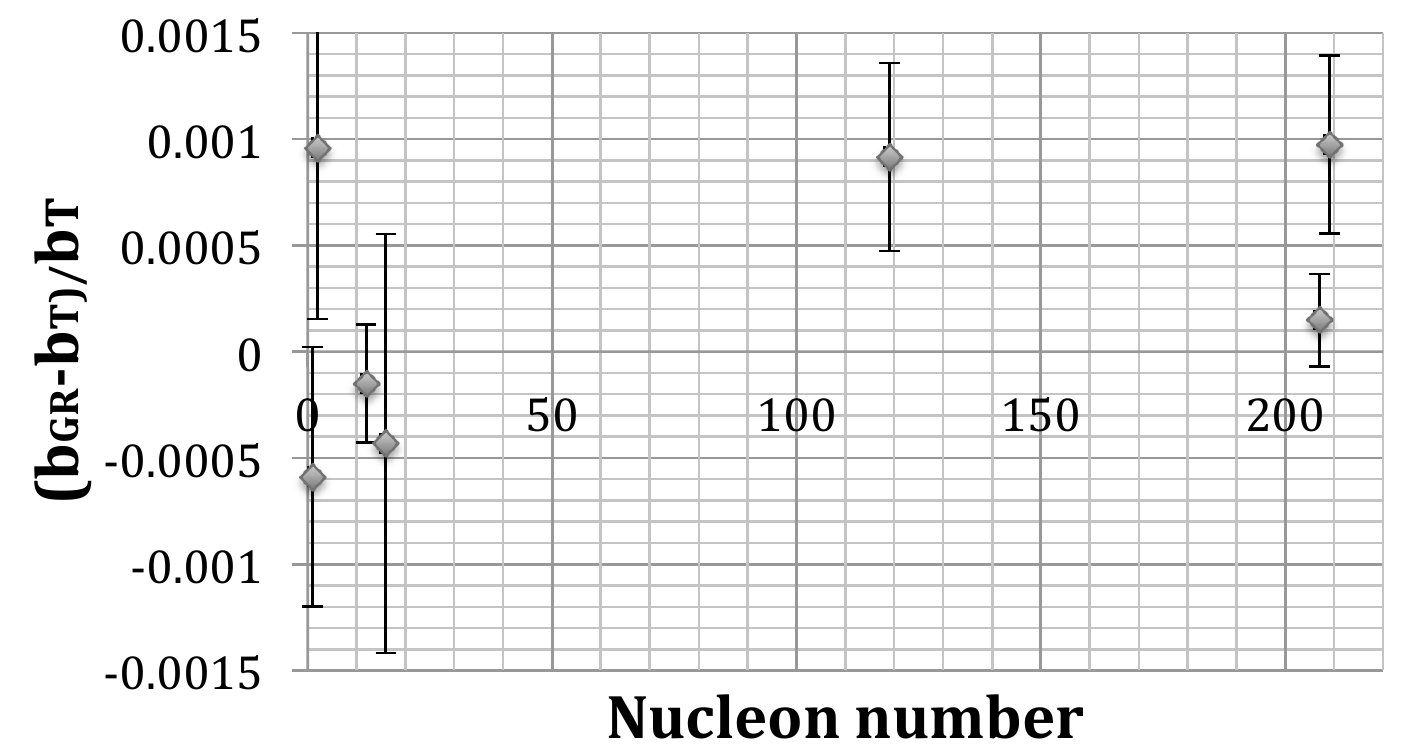}
    \caption{Fractional difference ${\Delta b}  \over {|b_{T}|}$ and between the coherent scattering amplitude $b_{GR}$ as measured by gravity reflectometry and $b_{T}$ as measured by neutron interferometry for the same media (hydrogen, deuterium, carbon, oxygen, tin, lead, and bismuth) along with uncertainties in $\Delta b  \over |b_{T}|$ plotted as a function of nucleon number A. The weighted mean of these 7 differences is $[2.2 \pm 1.4] \times 10^{-4}$.}
    \label{fig:data}
\end{figure}

With these upper bounds on the possible differences between the neutron scattering lengths measured by interferometry and gravity reflectometry and the corrections to the kinematic theory of neutron optics, we can test the internal consistency of the existing data. Table~\ref{tbl:lengths} presents the data set that we analyze to evaluate the internal consistency of the interferometry and gravity reflectometry data.  We used all the available high-precision data for isotopes that have been measured by both techniques. Fig.~\ref{fig:data} shows the difference ${\Delta b \over |b|}= {[b_\mathrm{GR}-b_{T}] \over |b_{T}|}$ as a function of A. The uncertainties for $\Delta b \over |b| $ for the 7 nuclei of interest come from the properly weighted sums of the results quoted in the references. The weighted mean of these 7 differences is $[2.2 \pm 1.4] \times 10^{-4}$. We conclude that this data is internally consistent at the $1.5 \sigma$ level. 

Future analyses which make use of new data and also other high-precision neutron cross section measurements which extend over a broader range of neutron energies and momentum transfers can make use of this feature to help determine an internally-consistent set of values for $b_{ne}$ and $b_{p}$ and constrain $\alpha$ and $\lambda$. An example of a successful global analysis of this type was presented long ago~\cite{Alexsejevs1998} for a broad set of neutron optics and scattering data for n-A scattering for $6<A<60$ using a S-matrix treatment of the potential and resonance scattering parameters, including a consistent treatment of effects from sub-threshold resonances.



\section{Possibilities for Future Improvements}
\label{sec:future}

One could extend this test of the internal consistency of these two methods of scattering length measurements if desired by measuring the coherent scattering lengths of a select set of nuclei and compounds using neutron interferometry to higher precision than they are known at present to yield a more sensitive comparison with the existing results from gravity reflectometry. We suggest measurements of carbon, carbon tetrachloride, gallium, and thallium, which can be compared to the already-measured gravity reflectometry results for Cl, Ga, and Th. The coherent scattering lengths for these elements have already been measured by neutron gravity reflectometry to precision near the $10^{-4}$ level needed for a sensitive consistency test. Aside from Th, which is poisonous in its pure form, the other materials can be obtained and handled easily in chemically pure form and with care can be formed into practical neutron interferometry targets. One can now obtain carbon of sufficient thickness, flatness, and density uniformity for precision neutron interferometry phase shift measurements in the form of artificial single crystal diamond plates, which are available with faces cut parallel to the crystal planes. CCl$_{4}$ is a liquid at room temperature which can be obtained in high chemical purity and can be placed in commercially-available rectangular quartz containers, which produce negligible small angle neutron scattering and whose internal thickness can be measured to high absolute accuracy using gauge blocks. Gallium is a liquid at 30\,C and therefore can be easily produced with a very uniform density and cast into a form with flat parallel sides. One would of course need to take care with to avoid and suppress bubbles in the liquid samples. This set of measurements would help cover the range of nucleon number $A$ more uniformly and could improve the precision of the consistency test described in this paper. The statistical accuracy of neutron interferometry phase shift measurements can approach 10 ppm as was shown in the measurements of the bound coherent neutron scattering length measurements on silicon~\cite{Ioffe1998}. 

One can in principle improve on the slow neutron beam reflectometry measurements of Koester {\it et al.} by another 1-2 orders of magnitude by instead using ultracold neutrons (UCN), which can be confined in material bottles at all angles of incidence and therefore correspond to neutrons with kinetic energies less than about $300$ neV. One could drop UCN onto a flat level sample surface, change its height, and measure the reflectivity curve as a function of $H$ as in the Koester approach. The GRANIT UCN spectrometer~\cite{Schmidt2009} nearing completion at the ILL/Grenoble, which is designed to conduct measurements on UCN gravitational bound states~\cite{Nesvizhevsky2002} and therefore already possesses a very flat, horizontal surface, could more sharply define the starting height of the neutrons than Koester et al.'s cold neutron apparatus. With the new superfluid-helium-based UCN source~\cite{Piegsa2014} installed to supply GRANIT, one could imagine measuring the neutron optical potential using the neutron reflectivity curve with statistical accuracy 1-2 orders of magnitude better than previous work. However the multiple scattering corrections to the optical potential relation to the scattering length are much larger for UCN than for cold neutrons and would need to be evaluated to higher precision than they are known now to be able to make full use of such data for our purposes. Good choices for the material to be used in such a measurement could be flat perfect crystals with known absolute densities at the ppm level such as silicon and germanium.




Another neutron measurement technique which could improve the sensitivity of the search for exotic Yukawa interactions is gravity resonance spectroscopy~\cite{Jenke2011, Abele2008}. This measurement technique creates coherent superpositions of bound states of neutrons formed in a potential from the Earth's gravity and a flat mirror, and one can drive and resolve resonance transitions using acoustic transducers in a vibrational version of Ramsey spectroscopy. The qBOUNCE apparatus has successfully conducted several measurements, including the proof of principle measurements demonstrating vibrational Rabi spectroscopy~\cite{Abele2010}, and has sought different types of exotic interactions~\cite{Ivanov2013, Jen14, Ivanov2016, Cronenberg2018, Klimchitskaya2019}. The eigenstate energies of the bouncing UCN would be shifted in the presence of an exotic Yukawa interactions sourced by the mirror material~\cite{Abele2003}. A new version of the qBOUNCE apparatus which is designed to implement vibrational Ramsey spectroscopy and has seen its first signal~\cite{Sedmik2019} has recently been commissioned. One can also consider employing a Lloyd's mirror interferometer for neutrons~\cite{Gudkov1993, Pokotolovskii2013a, Pokotolovskii2013b} as the interference between the forward-propagating amplitude and that reflected from the mirror in this type of interferometer can be sensitive to the exotic Yukawa phase shift from the mirror surface.

Dynamical diffraction in perfect crystals can measure $b(q)$ at larger values of $q$ of about an inverse Angstrom. However in this case many other effects must be corrected for, such as the contributions from the electromagnetic neutron-atom interaction proportional to $b_{ne}$ and due to the charge form factor of the electron cloud in the atom and also those from the Debye-Waller factor of the crystal, which at finite temperatures will need to include information on the phonon spectrum as well as possibly other material properties. The angular distribution of neutron scattering from noble gas atoms is sensitive to exotic Yukawa interactions through the $q$ dependence of the form factor in $b_{Y}(q)$ and has been used in two recent experiments which have improved the bounds on exotic Yukawa interactions with ranges near the Angstrom scale. The recent measurements using this approach which have improved the upper bounds on $\alpha$ for $\lambda$'s below $100$\,nm  can be improved. 

Having said all of this however: it is prudent also to emphasize some of the experimental difficulties and extra theoretical work which would need to be addressed in any such attempts to achieve the ${\Delta b \over |b|}=10^{-5}$ level of precision. One would certainly need to control and understand both the chemical purity, the knowledge of the isotopic composition of the materials, and possible density nonuniformities at an uncommon level of detail. In the case of perfect crystal neutron interferometry one must worry about possible corrections from geometric and  dynamical diffraction effects in the interferometer blades and one must control the external influences of the environment to a severe degree. The theory for the corrections to the kinematic theory of neutron optics discussed above would need to be improved, and as these theoretical corrections involve also knowledge of the internal structure and atom-atom correlations of the material it is likely that subsidiary measurements using neutron or maybe xray scattering would need to be performed on the samples as input to the theory corrections.   

\section{Conclusion}
\label{sec:conclusion}

We present the results of an internal consistency check on the experimental data from two different types of measurements of slow neutron scattering amplitudes on the same nuclei, neutron interferometry with $q^{2}=0$ and neutron gravity reflectometry with $q^{2}=8mV_{opt}$. We show that this consistency check is insensitive to possible corrections from electromagnetic and (possible) exotic Yukawa interactions in the narrow range of energies and momentum transfers accessed in these measurement methods. We show that the existing data is internally consistent at the $1.5 \sigma$ level. The fractional difference $\Delta b \over |b|$ averaged over the 7 elements where data exists on the same material from both measurement methods is $[2.2 \pm 1.4] \times 10^{-4}$. One must take into account some small corrections to the kinematic theory of neutron optics due to local field and multiple scattering effects to make this comparison. We view this exercise as an initial step in a future global analysis of neutron scattering data to bound possible exotic Yukawa interactions of the neutron. We outlined a number of ongoing measurement possibilities using slow neutrons, some now in progress, which could be used to improve the sensitivity of neutron-based exotic interaction searches. 

Although we considered possible neutron interactions of Yukawa form, other theories envision power-law interactions.  One could repeat the analysis presented in this paper for this case as well. Since power-law potentials fall off much more slowly than the damped exponential in the Yukawa potential, it is possible that power law potentials could introduce a larger difference between the neutron interferometry and neutron gravity reflectometry scattering amplitudes. In this case it is possible that the present data may already provide useful model constraints. Although we know of no simple analytical solutions for these cases, numerical analysis could be employed to solve for the modifications to the reflectivity and the accumulated phase shift in neutron interferometry. A comparison with the data presented in this paper as a function of nucleon number $A$ could then be used to constrain exotic interactions with power-law forms.     


\section{Acknowledgements}
\label{sec:ack}

All of the authors acknowledge support from US National Science Foundation grant PHY-1614545 and from the Indiana University Center for Spacetime Symmetries.
W. M. Snow acknowledges discussions with V. Nesvizhevsky on the future possibility of high-precision gravity reflectometry experiments using the GRANIT ultracold neutron spectrometer which occurred during a visit to Indiana University supported by a grant from the Gordon and Betty Moore Foundation.

\newpage
\onecolumngrid
\section{Appendix}
\noindent The reflection amplitude generalized to include Yukawa
interaction was presented by Taketani in terms of modified Bessel functions of the first kind as~\cite{Taketani2012}:\\\\
\begin{align}
  R &= -\frac{\phi_{o,+}(0)}{\phi_{o,-}(0)}
  \frac{\frac{d}{dz}\text{ln}\phi_{i,+}(z) - \frac{d}{dz}\text{ln}\phi_{o,+}(z)}{\frac{d}{dz}\text{ln}\phi_{i,+}(z) - \frac{d}{dz}\text{ln}\phi_{o,-}(z)}\bigg\rvert_{z=0}\\\nonumber
\end{align}

where as noted in the text $\phi_{o,+}$ ($\phi_{i,+}$) and $\phi_{o,-}$ ($\phi_{i,-}$) are the two independent solutions to the wave equation outside (inside) of the material. In the limit where the potential due to non-Newtonian gravity is much smaller than the kinetic energy, the following approximations can be made in this expression:\\\\

$\phi_{o,\pm}(0) \sim 1-x_o\pm i y_{o,a}$\\

$\frac{d}{dz}\text{ln}\phi_{o,\pm}(z)\bigg|_{z=0} \sim \pm i k_o(1 +
2x_o \mp i(y_{o,a}-y_{o,b} ))$\\

$\frac{d}{dz}\text{ln}\phi_{i,\pm}(z)\bigg|_{z=0} \sim \pm i k_i(1 + 2x_i \pm i(y_{i,a}-y_{i,b} ))$\\

\noindent where $k_{i,o}$ is the neutron wave vector for inside, outside the
material, and $\lambda$ is the interaction length for Yukawa-like gravity. The definitions for the other parameters are: \\

$\kappa_{o,i} = \frac{k_{o,i}}{k_g} = k_{o,i}\lambda$

$x_{o} = \frac{\kappa_{o}}{1+\kappa_{o}^2}\theta_0 =
\frac{k_{o}\lambda}{1+k_{o}^2\lambda^2}\left(-\frac{m_nV_g\lambda}{2k_o\hbar^2}\right)
= -\frac{m_nV_g\lambda^2}{(1+(k_o\lambda)^2)2\hbar^2}$\\

$y_{o,a} = \frac{\kappa_{o}^2}{1+\kappa_{o}^2}\theta_0 =\kappa_{o}x_o$\\

$y_{o,b} = \frac{1}{1+\kappa_{o}^2}\theta_0 =x_o/\kappa_o$\\

$x_{i} = \frac{\kappa_{i}}{1+\kappa_{i}^2}\theta_i =
\frac{k_{i}\lambda}{1+k_{i}^2\lambda^2}\left(\frac{m_nV_g\lambda}{2k_i\hbar^2}\right)
= \frac{m_nV_g\lambda^2}{(1+(k_i\lambda)^2)2\hbar^2}$\\

$y_{i,a} = \frac{\kappa_{i}^2}{1+\kappa_{i}^2}\theta_i =\kappa_{i}x_i$\\

$y_{i,b} = \frac{1}{1+\kappa_{i}^2}\theta_i =x_i/\kappa_i$\\

$\theta_o = -\frac{m_nV_g}{2k_ok_g\hbar^2} =
-\frac{m_nV_g\lambda}{2k_o\hbar^2}$\\

$\theta_i = -\frac{m_nV_g}{2k_ik_g\hbar^2} =
-\frac{m_nV_g\lambda}{2k_i\hbar^2}$\\

Substituting into the above expression for the reflectivity $R$ gives\\

\begin{align}
  R &\approx - \left(\frac{1-x_o+iy_{o,a}}{1-x_o-iy_{o,a}}\right)
  \frac{k_i(1+2x_i+i(y_{i,a}-y_{i,b})) - k_i(1+2x_i-i(y_{i,a}-y_{i,b}))}
  {k_i(1+2x_i+i(y_{i,a}-y_{i,b}))+k_i(1+2x_i+i(y_{i,a}-y_{i,b}))}\\\nonumber\\
&= - \left(\frac{1-x_o+iy_{o,a}}{1-x_o-iy_{o,a}}\right)
  \frac{[k_i(1+2x_i)-k_o(1+2x_o)]+i[k_i(y_{i,a}-y_{i,b})+k_o(y_{o,a}-y_{o,b})]}
{[k_i(1+2x_i)+k_o(1+2x_o)]+i[k_i(y_{i,a}-y_{i,b})+k_o(y_{o,a}-y_{o,b})]}.
\end{align}

\noindent The reflection probability is then given by \\ 

\begin{align}
 |R|^2 &= \frac{[k_i(1+2x_i)-k_o(1+2x_o)]^2+[k_i(y_{i,a}-y_{i,b})+k_o(y_{o,a}-y_{o,b})]^2}
         {[k_i(1+2x_i)+k_o(1+2x_o)]^2+[k_i(y_{i,a}-y_{i,b})+k_o(y_{o,a}-y_{o,b})]^2}\\\nonumber
         \label{eq:appR2}
\end{align}

\noindent Using $\frac{\hbar^2k_0^2}{2m_n} = m_n g H$ gives

\begin{align}
k_0 &= \sqrt{2m_n^2gH/\hbar^2}.\\\nonumber
\end{align}


\noindent At the critical height, $mgH_\text{c}^{\prime}$ = $V_\text{f}$ +
2$V_\text{g}$, so we can express $k_\text{i}$ in terms of $H_\text{c}^{\prime}$ and $H$:\\

\begin{align}
  \frac{\hbar^2k_i^2}{2m_n} &=   \frac{\hbar^2k_o^2}{2m_n} -
                              (V_\text{f}+2 V_\text{g})\\
                            &= m_ng(H-H_\text{c}^{\prime}), \nonumber 
\end{align}

\noindent where we use the primed critical height
$H_\text{c}^{\prime}$ to denote the usual critical height due to the
Fermi potential, $V_\text{f}$, with the addition of the Yukawa-like
gravitational potential $V_\text{g}$.\\

\begin{align}
  \Rightarrow  k_i &= \sqrt{2m_n^2g(H-H_\text{c}^{\prime})/\hbar^2}. 
\end{align}

\noindent The ratio is then given by 

\begin{align}
  \frac{k_i}{k_o} &= \sqrt{1-\frac{H_\text{c}^{\prime}}{H}},
\end{align}

\noindent and substituting into Eq.\,\ref{eq:appR2} gives 

\begin{align}
  |R|^2 &= \frac{
          \left[\sqrt{1-\frac{H_\text{c}^{\prime}}{H}}(1+2x_i)-(1+2x_o)\right]^2+\left[\sqrt{1-\frac{H_\text{c}^{\prime}}{H}}(y_{i,a}-y_{i,b})+(y_{o,a}-y_{o,b})\right]^2
          }
          {
          \left[\sqrt{1-\frac{H_\text{c}^{\prime}}{H}}(1+2x_i)+(1+2x_o)\right]^2+\left[\sqrt{1-\frac{H_\text{c}^{\prime}}{H}}(y_{i,a}-y_{i,b})+(y_{o,a}-y_{o,b})\right]^2
          }\\\nonumber\\
          &= \frac{
          \left[\sqrt{1-\frac{H_\text{c}^{\prime}}{H}}(1+2x_i)-(1+2x_o)\right]^2+\left[\sqrt{1-\frac{H_\text{c}^{\prime}}{H}}(\kappa_ix_i-x_i/\kappa_i)+(x_o\kappa_o-x_o/\kappa_o)\right]^2
          }
          {
          \left[\sqrt{1-\frac{H_\text{c}^{\prime}}{H}}(1+2x_i)+(1+2x_o)\right]^2+\left[\sqrt{1-\frac{H_\text{c}^{\prime}}{H}}(\kappa_ix_i-x_i/\kappa_i)+(x_o\kappa_o-x_o/\kappa_o)\right]^2
          }        \\\nonumber\\
          &= \frac{
          \left[\sqrt{1-\frac{H_\text{c}^{\prime}}{H}}(1+\frac{2\gamma}{1+\kappa_i^2})-(1-\frac{2\gamma}{1+\kappa_o^2})\right]^2+\gamma^2\left[\sqrt{1-\frac{H_\text{c}^{\prime}}{H}}(\frac{\kappa_i}{1+\kappa_i^2}-\frac{1}{\kappa_i(1+\kappa_i^2)})+(-\frac{\kappa_o}{1+\kappa_o^2}+\frac{1}{\kappa_o (1+\kappa_o^2)})\right]^2
            }
            {
          \left[\sqrt{1-\frac{H_\text{c}^{\prime}}{H}}(1+\frac{2\gamma}{1+\kappa_i^2})+(1-\frac{2\gamma}{1+\kappa_o^2})\right]^2+\gamma^2\left[\sqrt{1-\frac{H_\text{c}^{\prime}}{H}}(\frac{\kappa_i}{1+\kappa_i^2}-\frac{1}{\kappa_i(1+\kappa_i^2)})+(-\frac{\kappa_o}{1+\kappa_o^2}+\frac{1}{\kappa_o (1+\kappa_o^2)})\right]^2
          }\\\nonumber
\end{align}

\noindent where $\gamma = \frac{m_nV_g\lambda^2}{2\hbar^2}$ is a small
parameter proportional to the Yukawa-like gravitational potential
$V_g$, and $V_g$ is written in terms of a coupling parameter 
$\alpha_g$ as $V_g= Gm_n\pi\rho\,\alpha_g\lambda^2/2$.\\

\noindent Finally, since $\gamma \ll 1$ we can neglect terms proportional to it and rewrite the
reflectivity as\\

\begin{equation}
  |R|^2=
  \left[\frac{1-\sqrt{1-H_\text{c}^{\prime}/H}}{1+\sqrt{1-H_\text{c}^{\prime}/H}}\right]^2. 
  \label{eq:fres_mod}
\end{equation}

\end{document}